\documentclass[a4paper,11pt]{article}
\usepackage[utf8]{inputenc}
\usepackage[T1]{fontenc}
\usepackage[english]{babel}
\usepackage[width=160mm,height=247mm]{geometry}
\usepackage{amsmath, amsfonts, amssymb, bm}
\usepackage[pdftex]{graphicx}
\usepackage{lmodern, indentfirst, cite}
\usepackage[squaren]{SIunits}
\usepackage[textfont={footnotesize},labelfont={color=blue,bf,sf},labelsep=endash]{caption}
\usepackage[colorlinks=true,linkcolor=blue,citecolor=blue,urlcolor=blue]{hyperref}

\usepackage[affil-sl]{authblk}
\usepackage{hyperref}

\title{\textbf{\Large Thermal quantum coherence of two--qutrit Heisenberg XXZ model with Herring–Flicker coupling and Dzyaloshinskii-Moriya interaction under magnetic field}}
\author[a]{Brahim Adnane
}
\author[b]{Younes Moqine
}
\author[c]{Abdelhadi Belouad
}
\author[a]{El Bouâzzaoui Choubabi}
\author[a,c]{Rachid Houça\thanks{r.houca@uiz.ac.ma}}

\affil[a]{LPMC. Laboratory, Theoretical Physics Group, Faculty of Sciences, Choua\"ib Doukkali University, PO Box 20, 24000 El Jadida, Morocco}
\affil[b]{Research Laboratory of Physics and Engineers Sciences, Team of Applied Physics and New Technologies, Polydisciplinary Faculty (FP-BM), Sultan Moulay Slimane University, Béni Mellal, Morocco}
\affil[c]{LPTHE. Laboratory, Theoretical Physics and High Energy, Faculty of Sciences, Ibn Zohr University, PO Box 8106, Agadir, Morocco}
\date{}

\newcommand{\beq}{\begin{equation}}
\newcommand{\eeq}{\end{equation}}
\newcommand{\lb}{\label}

\begin{document}
\begin{titlepage}
	\newgeometry{width=175mm, height=247mm}
    \maketitle
    \thispagestyle{empty}
    \vspace{3cm}

\begin{abstract}
In this study, we use the concept of $l_1$-norm coherence to characterize the entanglement of a two--qutrit Heisenberg XXZ model for subject to a uniform magnetic field and z--axis Dzyaloshinskii--Moriya interaction with Herring-Flicker coupling. We show the temperature, magnetic field,  DM interaction, and distance of Herring-Flicker coupling all can control the entanglement. However, the state system becomes less entangled at high temperatures or strong magnetic fields and vice versa. Our findings also suggest that entanglement rises when the z--axis DM interaction increases. Moreover,  by setting the strengths coupling of the spin, we quickly recover the isotropic XY and XXX Heisenberg models. Finally, Herring-Flicker coupling affects the degree of entanglement. In fact, when Herring-Flicker coupling and temperature are at small values, the degree of entanglement is at its highest. Still, when Herring-Flicker coupling is at substantial values, the degree of entanglement tends to stabilize.
\end{abstract}

\vspace{2cm}



\end{titlepage}

\section{Introduction}
Entanglement is one of the most intriguing aspects of quantum physics \cite{Einstein}. It happens when particles interact in a manner that makes it impossible to characterize each particle’s quantum state separately. With implementations in quantum communication and teleportation \cite{Houca1,Bennett,Ekert}, quantum information processing \cite{Divincenzo}, quantum dense coding, and the application of various quantum protocols \cite{Grover,Shor,Khan}, quantum entanglement uses are a precious resource for research that cannot be done using conventional resources. Different physics fields have also successfully attained entanglement. For instance, quantum logic operations involving trapped ions \cite{Kielpinski}, nuclear spins of organic molecules \cite{Bogani}, semiconductor devices \cite{Chtchelkatchev}, and atomic chips \cite{Folman} have all been demonstrated to exhibit quantum communication patterns.

Finding a way to tell if a specific quantum system state is entangled or not, as well as selecting the optimal way to quantify the degree of entanglement, are significant challenges. In light of this, one of the most critical issues in the realm of quantum information is the quantization and characterization of the degree of entanglement. When the quantum system is in a pure state, the notion of entanglement is simpler and easier to comprehend. In contrast, the characterization of complete entanglement characteristics of mixed states is a challenging and unanswered mathematical problem. Several measures, such as negativity \cite{Peres,Horodecki,Zyczkowski,Vidal}, have been suggested to quantify entanglement.

Condensed matter systems’ quantum entanglement is a significant field, as is well known. Various studies on quantum entanglement have been accomplished on the thermal equilibrium states of spin chains subjected to an external magnetic field at a fixed temperature \cite{Amesen,WangX.G,ZhouL,ZhangG.F,Guo1}. Moreover, two-qubit quantum correlations with the Dzyaloshinskii--Moriya interaction (DM) receive much attention from researchers \cite{Tufarelli,YaoY1,Song23,Yao2,Liu23}. In addition, the Heisenberg model was used to study entanglement. A number of important works were produced, including the isotropic Heisenberg XX model \cite{Wang X G33,Wang X G34}, the XXX model \cite{Houca2,Nielsen,Arnesen}, the anisotropic Heisenberg XY model \cite{Kamta G L,Wang X G35}, the completely anisotropic Heisenberg XYZ model \cite{Zhou L36}, as well as some new research on the spin--1/2 \cite{William K}. Unfortunately, since measurements for higher spin systems are lacking, entanglement in spin--1 systems has gotten less attention. Vidal and Werner established a measure of entanglement called negativity, which may apply to higher spin systems \cite{Vidal G39,Schliemann40}. Using the notion of negative, Wang et al. arrived at analytical conclusions about entanglement in a spin--1 chain \cite{Sun44,Wang45}.

Now, we describe a few studies by various authors that include the coupling strength, $J$, as a function of location in various spin chain configurations. In reality, due to quantum fluctuations and noise, the coupling intensity in quantum systems may be regarded as a factor in practical scenarios regarding the distance separating quantum particles. This method could be utilized to improve quantum implementations. In $1988$ Haldane and Shastry \cite{hal,sha} presented the first effort to explore entanglement in the spin chain with long-range interactions, where the coupling strength is equal to the inverse of the distance square and follows the squared law. In a similar manner, B. Lin et al. \cite{lin} investigated the XXZ Heisenberg chain with long-range interactions, whereas M. XiaoSan et al. \cite{xia} investigated the XX Heisenberg model with Calogero Moser interactions.

Due to the significance of Herring-Flicker coupling (HF) in quantum information physics to determine the degree of intricacy between particles and motivated by previous works, we will investigate the thermal entanglement of spin--1 in a two-spin Heisenberg XXZ system with z-axis DM interaction, HF coupling distance, and in the presence of a uniform magnetic field. To the best of our knowledge, this study is the first to offer the first finding of thermal quantum correlation measured by negativity for two-spins-qutrit over the HF coupling distance.

This paper is organized as follows: Section {\color{blue}2} presents in--depth the theoretical model of our system exposed to DM interaction and a magnetic field with HF coupling. In addition, we establish the spectrum of the system at a specific temperature. Section {\color{blue}3} will address thermal quantum entanglement measured by $l_1$-norm coherence. Section {\color{blue}4} will be devoted to some exceptional cases, notably the isotropic XY and XXX models. Section {\color{blue}5} will include numerical studies to highlight the behavior of systems. Finally, the investigation is concluded with a summary of the results.
\section{Hamiltonian of the system}
In this study, we take into account a Heisenberg XXZ model with two--qutrit (spin 1), Herring-Flicker (HF) coupling, and z--axis DM interaction exposed to a uniform magnetic field. The Hamiltonian of the system is expressed by
\beq
\mathcal{H}=\mathcal{H}_H+\mathcal{H}_{DM}+\mathcal{H}_{Z}
\eeq
$\mathcal{H}_H$ corresponds to the XXZ Heisenberg chain, $\mathcal{H}_{DM}$ indicates the DM interaction, and $\mathcal{H}_{Z}$ refers to the Zeeman energy. Usually, the DM Hamiltonian $\mathcal{H}_{DM}$  can be represented as follows \cite{Moriya1,Moriya2,yones,brahim,houcaTel}
\begin{eqnarray}
\mathcal{H}_{DM} &=& \overrightarrow{D}. \left(\overrightarrow{\sigma}_1\times\overrightarrow{\sigma}_{2}\right) \\ \nonumber
 &=& D_x\left(\sigma_1^y\sigma_{2}^z-\sigma_1^z\sigma_{2}^y\right)+D_y\left(\sigma_1^x\sigma_{2}^z-\sigma_1^z\sigma_{2}^x\right)+
 D_z\left(\sigma_1^x\sigma_{2}^y-\sigma_1^y\sigma_{2}^x\right)
\end{eqnarray}
Where $\sigma^{x, y, z}$ denotes the Pauli matrices for a spin--1 and $D_{x, y, z}$ reflects the components of DM interaction. In the current studies, we limit ourselves to the case where the DM  interaction occurs along the z--axis. Then, our Hamiltonian has the form
\beq\lb{fr}
\mathcal{H}=J \left(\sigma_1^x\sigma_{2}^x+\sigma_1^y\sigma_{2}^y+\gamma\sigma_1^z\sigma_{2}^z\right)+D_z \left(\sigma_1^x\sigma_{2}^y-\sigma_1^y\sigma_{2}^x\right)+B \left(\sigma_{1}^z+\sigma_2^z\right)
\eeq
where $\sigma^m (m=x,y,z)$ denotes the spin--1 Pauli matrices given by
\beq
\sigma^x={1\over\sqrt{2}}\left(
           \begin{array}{ccc}
             0 & 1 & 0 \\
             1 & 0 & 1 \\
             0 & 1 & 0 \\
           \end{array}
                    \right) ,\quad \sigma^y={1\over\sqrt{2}}\left(
           \begin{array}{ccc}
             0 & -i & 0 \\
             i & 0 & -i \\
             0 & i & 0 \\
           \end{array}
         \right),\quad \sigma^z=\left(
           \begin{array}{ccc}
             1 & 0 & 0 \\
             0 & 0 & 0 \\
             0 & 0 & -1 \\
           \end{array}
         \right)
\eeq
and $D_z$ denotes the DM interaction along the z-axis. It should be noted that $J$ represents the coupling between the spin chains. Whenever the value of $J > 0$, the chain is antiferromagnetic; whenever the value of $J< 0$, the chain is ferromagnetic; $\gamma$ is the anisotropy parameter. Furthermore, we assume that the coupling $J$ is an HF coupling, i.e., $J(R)$, is defined by
\beq
J(R)=1.642\exp(-2R)R^{5\over2}+O(R^2\exp(-2R))
\eeq

\begin{figure}[!h]
  \centering
  \includegraphics[width=8 cm]{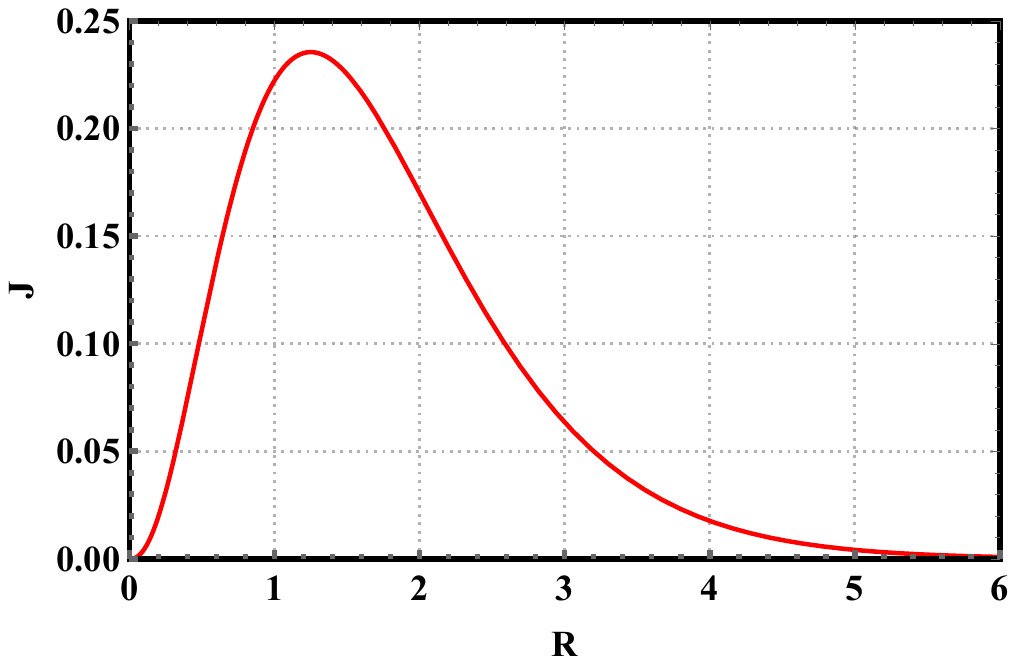}
  \caption{(Color online) The coupling $J$ of Herring-Flicker in terms of the distance $R$ between two spins.}\label{figure1}
\end{figure}
Where $R$ is HF coupling distance. The graph of the function $J$ in terms of $R$ is shown in Figure \ref{figure1}. From Fig. \ref{figure1}, it is evident that the coupling of HF is zero when the spins are far apart ($R> 6$) and has a maximum when they are closer ($R\simeq1.3$), implying that the spins become free at greater distances. As a result, in this study, we restrict ourselves to values on the margin where $J$ is non-zero $(0< R<6)$.

To evaluate the spectrum of the Hamiltonian  \eqref{fr} it is convenient to give the matrix form of $\mathcal{H}$ in the basis  $|-1,-1\rangle$, $|-1,0 \rangle$, $|-1,1 \rangle$, $|0,-1 \rangle$, $|0,0 \rangle$, $|0,1\rangle$, $|1,-1 \rangle$, $|1,0 \rangle$, $|1,1 \rangle$ as
\beq\lb{1}
\mathcal{H}=\left(
\begin{array}{ccccccccc}
 \gamma  J(R)+2 B & 0 & 0 & 0 & 0 & 0 & 0 & 0 & 0 \\
 0 & B & 0 & re^{i \theta }  & 0 & 0 & 0 & 0 & 0 \\
 0 & 0 & -\gamma  J(R) & 0 & re^{i \theta }  & 0 & 0 & 0 & 0 \\
 0 & re^{-i \theta }  & 0 & B & 0 & 0 & 0 & 0 & 0 \\
 0 & 0 & re^{-i \theta }  & 0 & 0 & 0 & re^{i \theta }  & 0 & 0 \\
 0 & 0 & 0 & 0 & 0 & -B & 0 & re^{i \theta }  & 0 \\
 0 & 0 & 0 & 0 & re^{-i \theta }  & 0 & -\gamma J(R) & 0 & 0 \\
 0 & 0 & 0 & 0 & 0 & re^{-i \theta }  & 0 & -B & 0 \\
 0 & 0 & 0 & 0 & 0 & 0 & 0 & 0 & \gamma  J(R)-2 B \\
\end{array}
\right)
\eeq
which the quantities $r$ and $\theta$ are defined by
\begin{eqnarray}
  r &=& \sqrt{D_z^2+J(R)^2} \\
  \theta &=&  \arctan\left(\frac{D_z}{J(R)}\right)
\end{eqnarray}
the eigenvalues as well as the eigenvectors of the equation \eqref{1} are given by
\begin{eqnarray}
\epsilon_{1,2} &=&B \pm r \lb{22}\\
\epsilon_{3,4} &=& \gamma J(R) \pm2 B\lb{222}\\
\epsilon_{5} &=& -\gamma J(R) \\
\epsilon_{6,7}&=&-B\pm r\\
\epsilon_{8,9}&=&\frac{r }{2}\chi_{2,1}
\end{eqnarray}
where $\chi_{1,2}=\frac{\sqrt{\gamma ^2 J(R)^2+8 r^2}\pm\gamma  J(R)}{r}$, and the related eigenvectors
\begin{eqnarray}\lb{psi}
  |\varphi_{1,2}\rangle &=&  \pm\frac{e^{i \theta }}{\sqrt{2}}|-1,0\rangle +\frac{1}{\sqrt{2}}|0,-1\rangle \\ \nonumber
  |\varphi_{3,4}\rangle &=& |\mp1,\mp1\rangle \\ \nonumber
  |\varphi_5\rangle &=& -\frac{e^{2 i \theta }}{\sqrt{2}}|-1,1\rangle+\frac{1}{\sqrt{2}}|1,-1\rangle \\ \nonumber
  |\varphi_{6,7}\rangle &=&\pm\frac{e^{i \theta }}{\sqrt{2}}|0,1\rangle+\frac{1}{\sqrt{2}}|1,0\rangle\\ \nonumber
  |\varphi_{8,9}\rangle &=&\frac{2 e^{2 i \theta }}{\sqrt{\chi _{1,2}^2+8}}|-1,1\rangle\pm\frac{e^{i \theta } \chi _{1,2}}{\sqrt{\chi _{1,2}^2+8}}|0,0\rangle+\frac{2}{\sqrt{\chi _{1,2}^2+8}}|1,-1\rangle
\end{eqnarray}
After obtaining the spectrum of our system, it is easy to calculate the thermal density, which is essential for performing measurements of the examined system's entanglement. To this end, the following section will be devoted to calculating the density matrix following the system's parameters.
\section{Thermal density matrix and negativity}
\subsection{Thermal density matrix}
After determining the system's spectrum, we shall seek the equation of the density matrix $\varrho(T)$, which will enable us to quantify the negativity at the thermal thermodynamic equilibrium at a given temperature $T$. According to the above-described framework, the system's state of thermal equilibrium might be expressed as follows
\beq
\varrho(T)={1\over\mathbb{Z}}e^{-\beta \mathcal{H}}
\eeq
such that the canonical partition function $\mathbb{Z}$ is written by
\beq
\mathbb{Z}=\sum_{i=1}^9 e^{-\beta \epsilon_i}
\eeq
such as $k_B$ is the constant of the Boltzmann, for convenience, it is considered as unity in the next. The temperature's inverse is represented by the parameter $\beta=1/T$. The spectrum of the Hamiltonian \eqref{1} allows expressing the thermal density $\varrho(T)$ as
\beq\lb{3}
\varrho(T)={1\over\mathbb{Z}}\sum_{k=1}^{4}e^{-\beta \epsilon_k}|\phi_k\rangle\langle\phi_k|
\eeq
Adding equations \eqref{22} and \eqref{psi} to equation \eqref{3}  yields the density matrix of the system in thermal equilibrium, which may be written on the previous basis as
\beq\lb{43}
\varrho(T)={1\over\mathbb{Z}}\left(
\begin{array}{ccccccccc}
 \rho _{11} & 0 & 0 & 0 & 0 & 0 & 0 & 0 & 0 \\
 0 & \rho _{22} & 0 & e^{i \theta } \rho _{24} & 0 & 0 & 0 & 0 & 0 \\
 0 & 0 & \rho _{33} & 0 & e^{i \theta } \rho _{35} & 0 & e^{2 i \theta } \rho _{37} & 0 & 0 \\
 0 & e^{-i \theta } \rho _{42} & 0 & \rho _{44} & 0 & 0 & 0 & 0 & 0 \\
 0 & 0 & e^{-i \theta } \rho _{53} & 0 & \rho _{55} & 0 & e^{i \theta } \rho _{57} & 0 & 0 \\
 0 & 0 & 0 & 0 & 0 & \rho _{66} & 0 & e^{i \theta } \rho _{68} & 0 \\
 0 & 0 & e^{-2 i \theta } \rho _{73} & 0 & e^{-i \theta } \rho _{75} & 0 & \rho _{77} & 0 & 0 \\
 0 & 0 & 0 & 0 & 0 & e^{-i \theta } \rho _{86} & 0 & \rho _{88} & 0 \\
 0 & 0 & 0 & 0 & 0 & 0 & 0 & 0 & \rho _{99} \\
\end{array}
\right)
\eeq
such that the density matrix elements are given by
\begin{eqnarray}\lb{23}
\varrho _{11}&=&e^{-\beta  (2 B+\gamma  J)}\\ \nonumber
\varrho _{22}&=&\varrho _{44}=e^{-\beta B} \cosh (\beta  r) \\ \nonumber
\varrho _{33}&=&\varrho _{77}=\frac{1}{2} e^{\beta  \gamma  J}+\frac{4 e^{-\frac{\beta  r \chi_2}{2} }}{\chi_1^2+8}+\frac{4 e^{\frac{\beta  r \chi_1}{2}}}{\chi_2^2+8} \\ \nonumber
\varrho _{55}&=&\frac{\chi _1^2 e^{-\frac{1}{2} \beta  r \chi _2}}{\chi _1^2+8}+\frac{\chi _2^2 e^{\frac{1}{2} \beta  r \chi _1}}{\chi _2^2+8}\\ \nonumber
\varrho _{37}&=&\varrho _{73}=\frac{1}{2} \left(-e^{\beta  \gamma  J}+\frac{8 e^{-\frac{\beta  r \chi_2}{2} }}{\chi_1^2+8}+\frac{8 e^{\frac{\beta  r \chi_1}{2}}}{\chi_2^2+8}\right) \\ \nonumber
\varrho _{35}&=&\varrho _{53}=\varrho _{57}=\varrho _{75}=-\frac{4 \left(e^{\frac{\beta  \gamma  J}{2}} \sinh \left(\frac{1}{4} \beta  r (\chi_1+\chi_2)\right)\right)}{\chi_1+\chi_2}\\ \nonumber
\varrho _{66}&=&\varrho _{88}=e^{\beta  B} \cosh (\beta  r)\\ \nonumber
\varrho _{24}&=&\varrho _{42}=-e^{-\beta B} \sinh (\beta  r) \\ \nonumber
\varrho _{68}&=&\varrho _{86}=-e^{\beta  B} \sinh (\beta  r)\\ \nonumber
\varrho _{99}&=&e^{\beta(2B-\gamma  J)} \nonumber
\end{eqnarray}
\subsection{ The $l_1$-norm coherence}
Under a specified reference basis  ${|i\rangle}$, a density $\varrho$ is told to be incoherent if the state is diagonal in this basis; in this case, we can write 
\begin{equation}
\varrho=\sum_i\rho_i|i\rangle\langle i|.
\end{equation}
Otherwise, the quantum state is coherent. For coherent states, the $l_1$-norm coherence and the entropy coherence are two typically used coherence measures \cite{Baumgratz}. The $l_1$-norm coherence of the quantum state can be expressed by 
\begin{equation}\lb{rf}
\varrho=\sum_{i,j}\rho_{i,j}|i\rangle\langle j|.
\end{equation}
is the sum of the magnitudes of all the off-diagonal entries:
\beq
\mathcal{C}_{l_1}(\varrho)=\sum_{j\neq j} |\rho_{ij}|.
\eeq
By replacing the Eqs. \eqref{43} and \eqref{23} into the Eq. \eqref{rf}, the literal expression of coherence is written by  
\begin{equation}\lb{coh}
\mathcal{C}_{l_1}=\frac{2}{\mathbb{Z}} \left(2 \cosh (\beta  B) | \sinh (r \beta )| +\left| -\frac{1}{2} e^{J \beta  \gamma }+\frac{4 e^{\frac{1}{2} r \beta  \chi _1}}{\chi _2^2+8}+\frac{4 e^{-\frac{r \chi _2}{2}}}{\chi _1^2+8}\right| +4 \left| \frac{e^{-\frac{1}{2} r \beta  \chi _2} \chi _1}{\chi _1^2+8}-\frac{e^{\frac{r \beta  \text{$\chi $1}}{2}} \chi _2}{\chi _2^2+8}\right| \right).
\end{equation}
The previous expression shows that the coherence depends on the temperature, DM coupling, anisotropy and the coupling parameter between the spin. Indeed, the impact of these quantities will be studied in detail in the following sections.
\section{Special cases}
Before starting the numerical study, we discuss two particular cases. Using the analytical formula in the equation \eqref{coh} and adjusting some quantities, such as the spin couplings $J$ and $\gamma$, we give a detailed study for the general XXZ Heisenberg model. Then, we shall consider the isotropic XY model $(\gamma= 0) $ and the XXX model ($\gamma=1$). Now, we arrive to analyze the amount of quantum entanglement via the coherence $\mathcal{C}_{l_1}$ in spin systems in the presence of a uniform magnetic field with DM  interaction.
	\subsection{Isotropic XY model}
To start, we consider {\color{red}the isotropic XY  model} corresponding to the case $(\gamma= 0) $, which implies that  $r= J$ and $\chi_1=\chi_2=2 \sqrt{2}$ with $D_z=B=0$ , which indicates that $\mathbb{Z}=4 \cosh (\beta  J)+2 \cosh \left(\sqrt{2} \beta  J\right)+3$. In this case, to obtain the expression of coherence, replace the quantities below in Eq. \eqref{coh} to get
	\begin{eqnarray}\label{fdf}
	\mathcal{C}_{l_1}=\frac{ 2\sqrt{2}  \left| \sinh(\sqrt{2}\beta J)\right| +4 | \sinh ( \beta J )| +\cosh \left(\sqrt{2} \beta  J\right)-1}{4 \cosh (\beta  J)+2 \cosh \left(\sqrt{2} \beta  J\right)+3}
	\end{eqnarray}
	\begin{figure}[!h]
		\centering{\includegraphics[width=9.6cm]{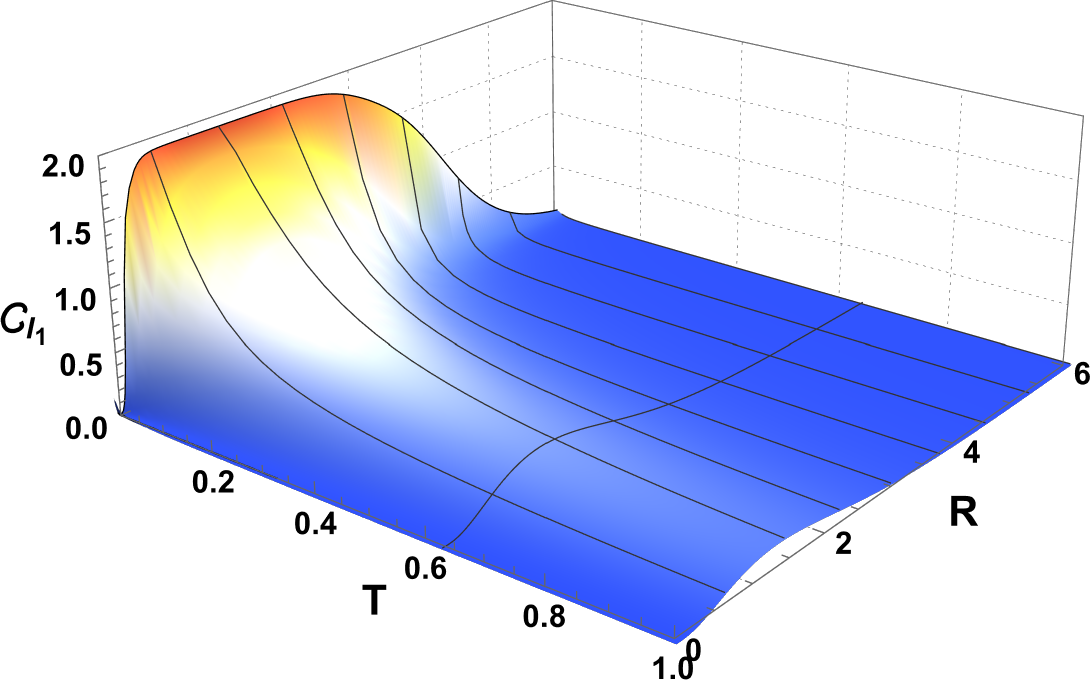}}
		\caption{(Color online) The coherence versus  $T$ and $R$ for isotropic XY model.}\label{XX}
	\end{figure}
	
From figure \ref{XX}, we notice that for extremely low temperatures and small values of $R$, coherence is maximal and that it declines monotonically with temperature until a threshold temperature $T_c$ is reached, which is the solution of $ 2\sqrt{2}  \left| \sinh(\sqrt{2}J/T_c)\right|+4 | \sinh (J/T_c )| +\cosh \left(\sqrt{2} J/T_c\right)= 1$. We also can conclude that systems that have significant coupling $J$ exhibit more remarkable coherence at $T=0$ and $0<R<3$.  Moreover, the coherence is the same whether the system is ferromagnetic ($J <0$) or antiferromagnetic ($J > 0)$. 
	\subsection{Isotropic XXX model}
	To recover the isotropic XXX  model corresponding to the case ($r=  J$), which implies that $\chi _1=2 \chi _2=4$  with $D_z=B=0$, which indicates that $\mathbb{Z}=5 e^{\beta  (-J)}+3 e^{\beta  J}+e^{2 \beta  J}$. In this case, to get the expression of concurrence, we replace the quantities below in Eq. \eqref{coh} to obtain
	\begin{eqnarray}\lb{aga}
		\mathcal{C}_{l_1}=\frac{1}{3} \left(\frac{4 \left| -1+e^{3 J \beta }\right| +12 e^{\beta  J} | \sinh (J \beta )| -9 \left(e^{2 \beta  J}+1\right)}{3 e^{2 \beta  J}+e^{3 \beta  J}+5}+2\right)
	\end{eqnarray}
	\begin{figure}[!h]
	\centering{\includegraphics[width=9.6cm]{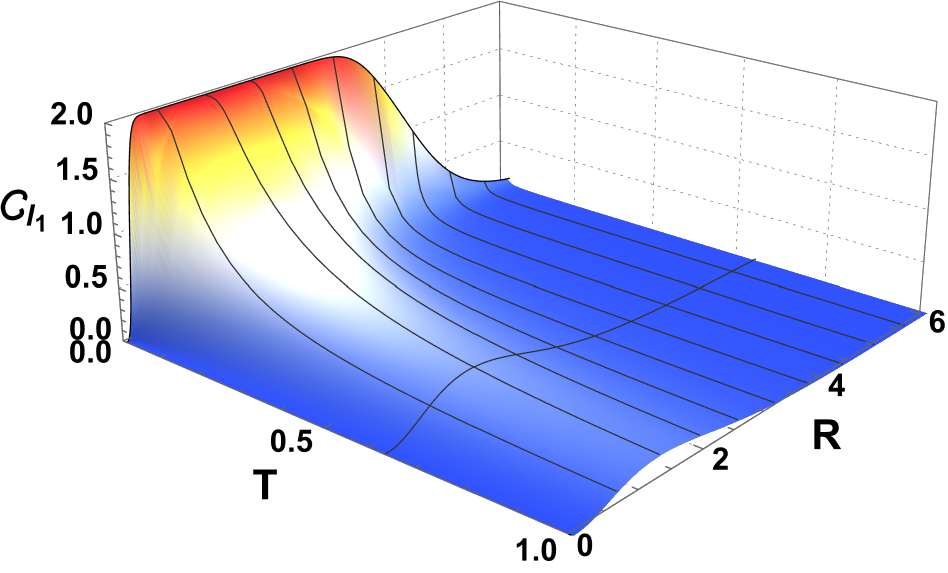}}
	\caption{(Color online) The coherence versus  $T$ and $R$ for isotropic XXX model..}\label{XXX}
\end{figure}
Figure \ref{XXX} shows the coherence of two qutrit XXX Heisenberg model. From Fig. \ref{XXX}, we observe the same remarks as the previous figure. The only difference is that the maximum value expected by the coherence at high temperatures and for small values of $R$ is greater than in the XY Heisenberg model. \\
We currently have all we need to investigate the behavior of our system. To do this, we will devote the following part to a numerical analysis of the coherence $\mathcal{C}_{l_1}$, as detailed below, to demonstrate the full performance of the proposed system. Then, we presented some graphs depending on the system's parameters under consideration, like temperature $T$, HF coupling distance $R$,  DM interaction $D_z$, and uniform magnetic field $B$. In addition, we will continue to examine this in more detail in order to draw conclusions.
\section{Numerical results}
This section will quantitatively investigate several elements of entanglement in a two--spin--qutrit Heisenberg XXZ chain with z--axis of DM interaction in terms of the HF coupling distance $R$ for various temperatures and magnetic field values. Firstly, we will explore coherence $\mathcal{C}_{l_1}$ as a function of temperature $T$ for different $R$ and magnetic field values by fixing the z--axis DM interaction such as $D_z=0.5$. Secondly, for a given temperature $T$, coupling distance $R$, and magnetic field $B$, we display the coherence $\mathcal{C}_{l_1}$ as a function of z--axis DM interaction. Finally, we will adjust the z--axis  DM interaction for the value $D_z=1$ to plot the coherence $\mathcal{C}_{l_1}$  as a function of the HF coupling distance $R$, temperature $T$, and magnetic field $B$.

\begin{figure}[!h]
  \centering{\includegraphics[width=16.5cm]{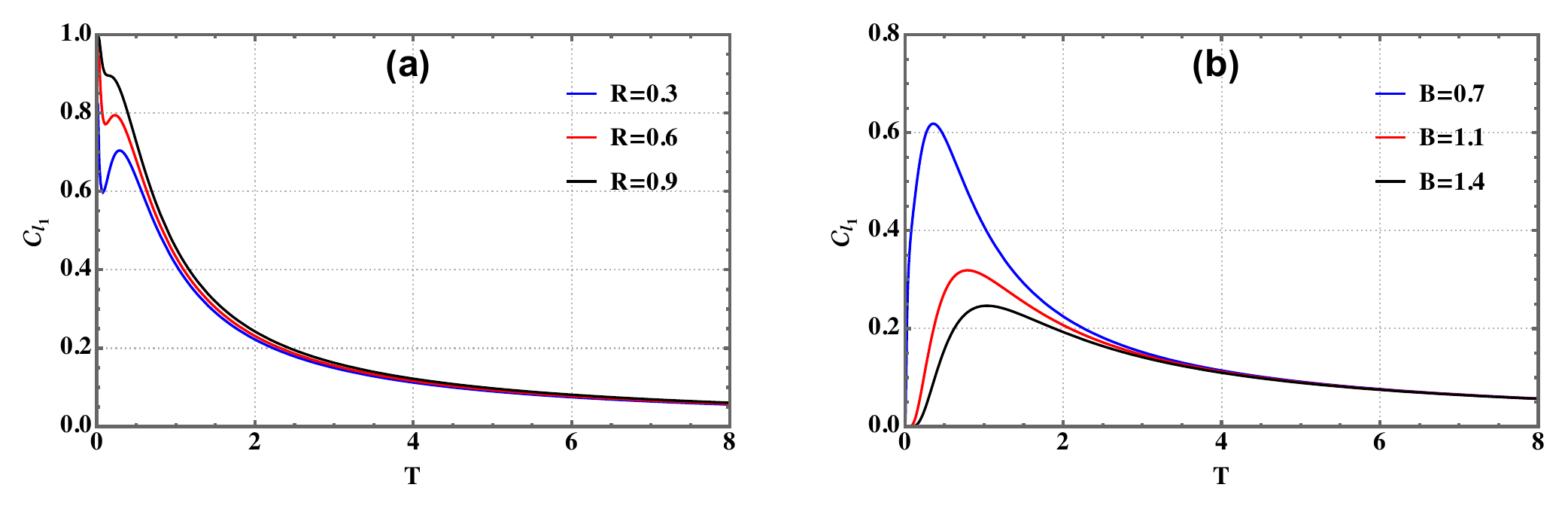}}
  \caption{(Color online) (a) The coherence versus  $T$ for different $R$ for $B=D_z=0.5$. (b) The cohernce versus $T$ for different $B$ for the fixed values $D_z=R=0.5$.}\label{f1}
\end{figure}
In figure (\ref{f1}), we plot the coherence $\mathcal{C}_{l_1}$ as a function of temperature for different HF coupling distances $R$ (Fig. \ref{f1}(a)) and magnetic field $B$ (Fig. \ref{f1}(b)). In Fig. \ref{f1}(a), the coherence is plotted as a function of the temperature $T$ for $R = 0.3, 0.6, 0.9$ and the other parameters are set as $D_z=B = 0.5$. At high temperatures, the coherence is cancelled, which means that the state's system becomes separable. Moreover, for a fixed value of $T$, the coherence increases with increasing HF coupling distance $R$. From Fig. \ref{f1}(b) when $B = 0$, the minimally entangled state  $|\varphi_3\rangle$ or $|\varphi_4\rangle$  is the ground state with eigenvalue $\gamma J(R)$ with the minimum entanglement, i.e., $\mathcal{C}_{l_1}= 0$. Furthermore, when $T$ rises, the coherence lowers owing to the mixture of other states with the maximally entangled state. As a result, as $T$ grows, the maximally entangled states mix with the unentangled states $|\varphi_3\rangle$ or $|\varphi_4\rangle$, hence increasing entanglement. Coherence decreases with rising magnetic field $B$ for a given value of $T$ until it reaches zero at higher temperatures.

\begin{figure}[!h]
  \centering
  \includegraphics[width=15.5cm]{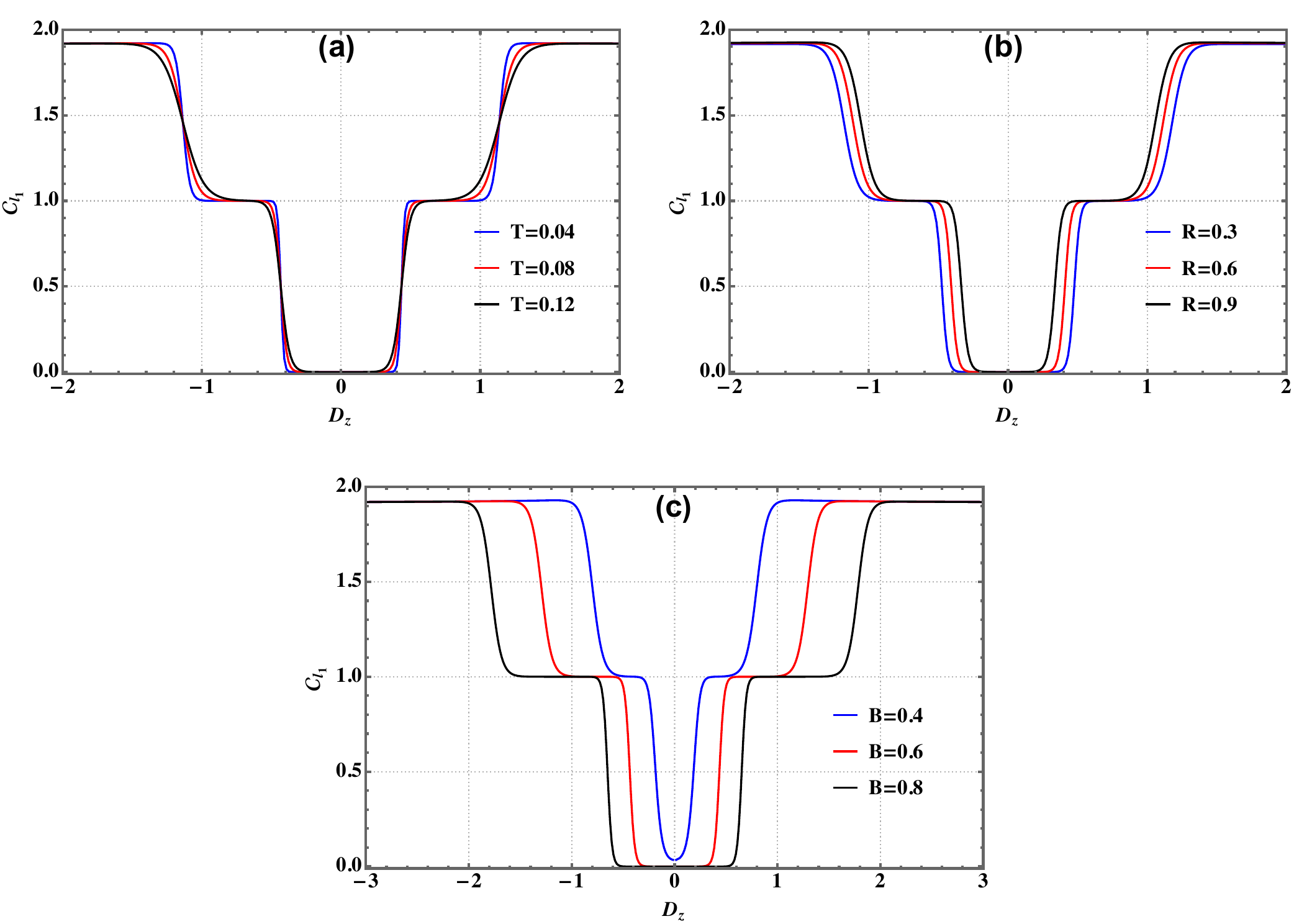}
  \caption{(Color online) The coherence versus z-axis DM interaction. (a) for different $T$ ($R=B=0.5$), (b) for different $R$ ($T=0.02$ and $B=0.5$), (c) for different $B$ ($T=0.02$ and $R=0.5$).}\label{f2}
\end{figure}
Figure \ref{f2} shows the coherence virsus  $D_z$ with different parameters ($T$, $R$ and $B$). The first remark is that the coherence is symmetrical compared to $D_z=0$. For this, we will interpret only the positive values of $D_z$. The coherence remains null for a certain value of $D_z$ lower than the critical value $D_z^c$, this last varying according to the parameters $T$, $R$ or $B$. However, the rise of the quantities $T$ and $R$ decreases the critical value of $D_z^c$ (Figs. \ref{f2}(a) and (b)). On the other hand, when the magnetic field increases, the $D_{z}^{c}$ critical value increases (Fig. \ref{f2}(c)). In Fig. \ref{f2}(b) and for the interval $D_z^c<D_z<1$, we notice the same behavior as in Fig. \ref{f2}(a).  Finally, when the DM interaction parameter $D_z$ is very high, the coherence tends to an identical maximum fixed value in all three figures, which means that the state's system is maximally entangled.

\begin{figure}[!h]
  \centering
  \includegraphics[width=15.5cm]{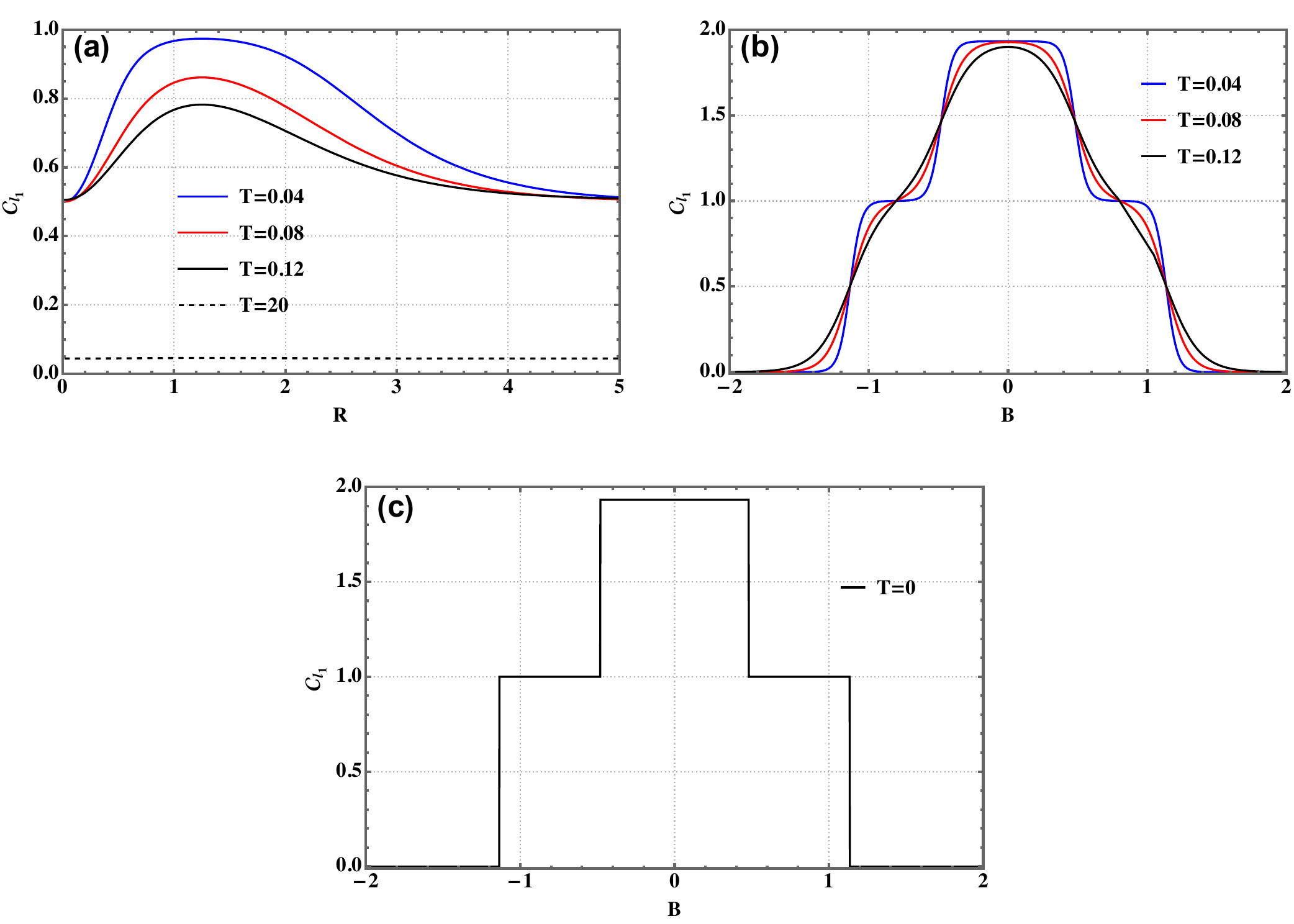}
  \caption{(Color online) (a): The coherence in terms of $R$  ($B=D_z=1$), and (b,c): versus $B$ ($R=D_z=1$) for different value of $T$.}\label{f3}
\end{figure}
In Fig. \ref{f3}, we display the coherence $\mathcal{C}_{l_1}$ as a function of the HF coupling distance parameter $R$ (Fig. \ref{f3}(a)), the uniform magnetic field $B$ (Fig. \ref{f3}(b)) at different temperature values $(0.04, 0.08, 0.12$ and  $20)$ and the uniform magnetic field $B$ at $T = 0$ (Fig. \ref{f3}(c)). In Fig. \ref{f3}(a), we see that the coherence increases when the parameter $R$ increases in the XXZ system with DM interaction ($D_z = 1$) and in the presence of a magnetic field ($B = 1$). When the HF coupling distance parameter $R$ is large, the coherence tends to be constant even as the temperature rises. Furthermore, it is clear that at high  temperature, the entanglement vanishes even if $R$ varied, which is explained by the dominance of the temperature (black dashed curve). From Figures \ref{f3}(b) and \ref{f3}(c), we see that there is evidence of phase transition at low temperature by increasing the magnetic field $B$. When $B$ is small, the entanglement is initially at its maximum. When the uniform magnetic field $B$ is greater, the entanglement decreases until it takes on a null value. At $T = 0$, we can also see that the entanglement vanishes when $B$ crosses a critical point $B_c$. However, the role of magnetic fields is mainly to reduce entanglement.
\section{summary and perspectives}
In this paper, we have explored thermal quantum entanglements by using the $l_1$-norm coherence concept to discuss the entanglement of a two--qutrit Heisenberg XXZ chain subjected to the uniform magnetic field and z--axis DM interaction with the distance of the HF coupling. The Hamiltonian model is described, the spectrum entanglement has been determined through mathematical calculations, and the thermal state at a finite temperature is mentioned explicitly. The numerical behavior of the $l_1$-norm coherence-measured entanglements in our study has been examined in terms of temperature, z--axis DM interaction, HF distance coupling, and uniform magnetic field. The ground state entanglement at zero temperature has been analyzed. We discovered that coherence declines monotonically with increasing temperature, and it is also clear that the magnetic field eliminates entanglement. Furthermore, We have also looked into the z-axis DM interaction parameter $D_z$, which can be raised to enhance entanglement. Finally, the effect of the HF coupling distance $R$ on quantum entanglement in spin systems is explored, i.e., decreasing the temperature can improve the entanglement of the system states. In addition, we have studied various exceptional cases, such as the isotropic XY and XXX Heisenberg models.

Still, some intriguing questions have to be addressed. Can we employ the investigated system to study thermal teleportation and dense coding or dynamic behaviors and demonstrate the fundamental aspects of quantum entanglement at a finite time? A related question appeared, what about other correlation measurements to check? These issues and associated questions are under consideration.
%
%
%
%

\end{document}